\documentclass[letterpaper,journal]{IEEEtran}

\usepackage{cite}
\usepackage{amsmath,amssymb,amsfonts}
\usepackage{graphicx}
\usepackage{booktabs}
\usepackage{tabularx}
\usepackage{array}
\usepackage{flushend}

\newcommand{\vect}[1]{\boldsymbol{#1}}
\newcommand{\E}{\mathbb{E}}
\newcommand{\clip}{\operatorname{clip}}

\newcommand{\resultfigure}[2]{%
  \IfFileExists{#1}{%
    \includegraphics[width=0.88\textwidth]{#1}%
  }{%
    \PackageWarning{TVTC}{Missing figure file: #1}%
    \fbox{\parbox[c][1.80in][c]{0.86\textwidth}{%
      \centering\small #2\\[4pt]
      \footnotesize Figure file missing: \texttt{\detokenize{#1}}}}%
  }%
}

\begin{document}

\title{Fairness-Guided DQN-Based TXOP Control for NR-U/Wi-Fi Coexistence}

\author{Po-Heng Chou,~\IEEEmembership{Member,~IEEE},
        Yi-Fang Yu,
        Shou-Yu Chen,
        and Chiapin Wang,~\IEEEmembership{Senior Member,~IEEE}%
\thanks{This work was supported in part by the National Science and Technology Council (NSTC) of Taiwan under Grants 113-2926-I-001-502-G and 114-2221-E-003-033. (Corresponding author: Chiapin Wang.)}%
\thanks{Po-Heng Chou is with the Research Center for Information Technology Innovation, Academia Sinica, Taipei 11529, Taiwan (e-mail: d00942015@ntu.edu.tw).}%
\thanks{Yi-Fang Yu, Shou-Yu Chen, and Chiapin Wang are with the Department of Electrical Engineering, National Taiwan Normal University, Taipei 10610, Taiwan (e-mail: nowonder201@gmail.com; 61475054h@ntnu.edu.tw; chiapin@ntnu.edu.tw).}\vspace{-0.4in}}

\maketitle

\begin{abstract}
New Radio operating in unlicensed spectrum (NR-U) and Wi-Fi employ different contention parameters and channel-occupancy durations, which can produce severe coexistence imbalance. This correspondence develops a fairness-guided deep Q-network (DQN) that adjusts the NR-U transmission opportunity (TXOP) using the observed per-node throughput ratio. Two reward designs, Q1 and Q2, expose complementary operating regimes. Q1 promotes strict per-node throughput balance, whereas Q2 admits a broader inter-radio-access-technology fairness region and rewards throughput improvement in an intermediate fairness band. The policies are compared with the default Third Generation Partnership
Project (3GPP) listen-before-talk (LBT), an adapted multi-armed-bandit
method, a utility-fairness-based tabular Q-learning comparator, and a default-referenced fixed-TXOP baseline. Experiments across four priority classes and asymmetric populations report aggregate throughput, user-level Jain fairness, and concave throughput utility over 20 independent trials. The results characterize the cost of strict fairness, efficiency recovery under moderate relaxation, and transfer from symmetric to asymmetric deployments.
\end{abstract}

\begin{IEEEkeywords}
Deep Q-network (DQN), NR-U, spectrum coexistence, transmission opportunity (TXOP), Wi-Fi.
\end{IEEEkeywords}

\section{Introduction}
\IEEEPARstart{N}{ew} Radio operating in unlicensed spectrum (NR-U) shares contention-based channels with Wi-Fi. Transmission opportunity (TXOP) control is central to their fair coexistence. NR-U employs the Third Generation Partnership Project (3GPP) listen-before-talk (LBT) mechanism, whereas Wi-Fi uses carrier-sense multiple access with collision avoidance (CSMA/CA). Their contention parameters and allowable channel-occupancy durations differ. An NR-U node can consequently retain the channel longer after successful access and reduce Wi-Fi access opportunities~\cite{3GPP_TS_37213,Bianchi2000JSAC,Cano2015LTEUCoexistence,Kakkad2023TVT}. TXOP adaptation directly controls this fairness--efficiency tradeoff. Such coexistence control is relevant to vehicular and mobile broadband deployments where NR-U and Wi-Fi share unlicensed spectrum.

Existing solutions address complementary coexistence-control problems. Analytical methods optimize fixed access parameters under a prescribed fairness definition~\cite{Luo2022ICC,Kakkad2023TVT,Alhulayil2025Access}, but their operating points depend on the node population and traffic model~\cite{Yingqi_2023}. Markov and learning-based studies also investigate buffering-aware licensed-assisted access, heterogeneous unlicensed-band allocation, and licensed/unlicensed band selection in device-to-device and sidelink networks~\cite{Chou_2019,Chou_2020,Chou_2024,Chou_2025}. These works optimize band allocation rather than post-access NR-U TXOP. Other learning-based schemes adapt access or scheduling from observations~\cite{Fasihi2024VTC,Zhou2022ICC,Ye2024TWC,Liu2023JSAC}. The adapted multi-armed-bandit comparator based on~\cite{Bajracharya2023TVT} learns candidate-decision rewards without conditioning on a coexistence state. A single prescribed reward also does not expose how different fairness preferences change TXOP behavior.

These prior studies leave three distinctions that are central to TXOP-only coexistence control. Band selection or scheduling does not directly control post-access channel-holding time, while a state-independent arm-selection rule cannot condition its decision on the same measured coexistence state as a value-based policy. Moreover, fairness between the average per-node throughputs of the two radio access technologies (RATs) can differ from fairness across all users when node populations are asymmetric. These differences motivate a compact state-aware formulation in which the reward preference and the reported evaluation metrics remain explicit.

This correspondence develops a fairness-guided deep Q-network (DQN) for state-aware TXOP control. Q1 uses the per-node throughput ratio to promote strict balance between the two RATs. Q2 uses an inter-RAT Jain index and rewards throughput improvement within an intermediate fairness band. For Q1 and Q2, concave throughput utility is used only as a common post-training evaluation metric and is not included in either reward.

The contributions are twofold. First, we formulate TXOP adaptation as a compact Markov decision process with a scalar throughput-ratio state and two explicit fairness-guided reward mappings. Second, we evaluate the policies across priority classes and asymmetric populations. Default 3GPP LBT, an adapted bandit~\cite{Bajracharya2023TVT}, a utility-fairness-based tabular Q-learning comparator, and a default-referenced fixed-TXOP baseline are compared using aggregate throughput, user-level Jain fairness, utility, and 95\% confidence intervals (CIs). The comparison identifies operating-point tradeoffs rather than claiming uniform superiority.

%\vspace{-0.2in}
\section{System Model and Metrics}
Consider one unlicensed channel shared by $N_{\mathrm{NR}}$ NR-U nodes and $N_{\mathrm{WF}}$ Wi-Fi nodes. These positive integers denote the respective populations. All nodes are saturated and belong to one collision domain. Nodes within each RAT use identical parameters. Let $k\in\{\mathrm{NR},\mathrm{WF}\}$ index a RAT, with $\mathrm{NR}$ and $\mathrm{WF}$ denoting NR-U and Wi-Fi; $N_k$ is its node count, and $\ell\ne k$ denotes the other RAT. Let $\tau_k$ be one RAT-$k$ node's transmission probability in a generic backoff slot. The probability that RAT $k$ is active is
$P_{\mathrm{tr},k}=1-(1-\tau_k)^{N_k}$, while the conditional probability
that exactly one node transmits given RAT activity is
$P_{\mathrm{s},k}=N_k\tau_k(1-\tau_k)^{N_k-1}/P_{\mathrm{tr},k}$.
The conditional collision probability observed by a RAT-$k$ node is
$p_k=1-(1-\tau_k)^{N_k-1}(1-\tau_\ell)^{N_\ell}$, where $\ell\ne k$.
Together with the adopted backoff functions
$\tau_k=f_k(p_k;W_{0,k},m_k)$, these relations form a fixed-point
system~\cite{3GPP_TS_37213,Bianchi2000JSAC}.
Here, $p_k$ is RAT $k$'s conditional collision probability, $W_{0,k}$ its initial contention window (CW) size, $m_k$ its maximum backoff stage, and $f_k(\cdot)$ its protocol-specific attempt-probability function. The system is solved numerically for each priority class and population before TXOP control. TXOP does not enter this fixed-point system; it changes occupancy durations rather than attempt probabilities.

Let $t=0,1,\ldots$ index control steps and $T_t$ be the NR-U TXOP at step $t$. Let $D_{\mathrm{NR}}=500~\mu\mathrm{s}$ be the fixed NR-U slot/alignment duration. The successful-transmission and intra-RAT collision durations are $T_{\mathrm{s},k}$ and $T_{\mathrm{c},k}$, respectively. The environment uses $T_{\mathrm{s,NR}}=T_{\mathrm{c,NR}}=T_t+D_{\mathrm{NR}}$ and fixed Wi-Fi durations $T_{\mathrm{s,WF}}$ and $T_{\mathrm{c,WF}}$. The mean generic-slot duration $\bar T_t$ is
\begin{align}
 \bar T_t={}&(1-P_{\mathrm{tr,NR}})(1-P_{\mathrm{tr,WF}})\sigma \nonumber\\
 &+P_{\mathrm{tr,NR}}P_{\mathrm{s,NR}}(1-P_{\mathrm{tr,WF}})T_{\mathrm{s,NR}}\nonumber\\
 &+P_{\mathrm{tr,WF}}P_{\mathrm{s,WF}}(1-P_{\mathrm{tr,NR}})T_{\mathrm{s,WF}}\nonumber\\
 &+P_{\mathrm{tr,NR}}(1-P_{\mathrm{s,NR}})(1-P_{\mathrm{tr,WF}})T_{\mathrm{c,NR}}\nonumber\\
 &+P_{\mathrm{tr,WF}}(1-P_{\mathrm{s,WF}})(1-P_{\mathrm{tr,NR}})T_{\mathrm{c,WF}}\nonumber\\
 &+P_{\mathrm{tr,NR}}P_{\mathrm{tr,WF}}T_{\mathrm{c,mix}},\label{eq:slot}
\end{align}
where $\sigma$ is the idle-slot duration and $T_{\mathrm{c,mix}}=T_{\mathrm{c,NR}}=T_t+D_{\mathrm{NR}}$ is the implemented simultaneous NR-U/Wi-Fi collision duration. Let $B_{k,t}$ denote the payload bits delivered by a successful RAT-$k$
transmission. Its aggregate throughput is
$\Gamma_{k,t}=P_{\mathrm{tr},k}P_{\mathrm{s},k}
(1-P_{\mathrm{tr},\ell})B_{k,t}/\bar T_t$, where $\ell\ne k$.
The average per-node throughput is $\bar\Gamma_{k,t}=\Gamma_{k,t}/N_k$, and the total throughput is $\Gamma_{\mathrm{sum},t}=\Gamma_{\mathrm{NR},t}+\Gamma_{\mathrm{WF},t}$.

The reward and evaluation use different fairness weightings. Q2 uses the inter-RAT Jain fairness index $J_t^{\mathrm{RAT}}$:
\begin{equation}
 J_t^{\mathrm{RAT}}=
 \frac{(\bar\Gamma_{\mathrm{NR},t}+\bar\Gamma_{\mathrm{WF},t})^2}
 {2(\bar\Gamma_{\mathrm{NR},t}^2+\bar\Gamma_{\mathrm{WF},t}^2)}.
 \label{eq:jain-rat}
\end{equation}
Under within-RAT throughput symmetry, the user-level Jain fairness index $J_t^{\mathrm{usr}}$ is
\begin{equation}
 J_t^{\mathrm{usr}}=
 \frac{(\Gamma_{\mathrm{NR},t}+\Gamma_{\mathrm{WF},t})^2}
 {(N_{\mathrm{NR}}+N_{\mathrm{WF}})
 (\Gamma_{\mathrm{NR},t}^2/N_{\mathrm{NR}}+
  \Gamma_{\mathrm{WF},t}^2/N_{\mathrm{WF}})}.
 \label{eq:jain-user}
\end{equation}
The former weights the two RAT-average per-node throughputs equally; the latter weights them by their populations. They coincide for equal populations but generally differ for unequal populations.

For Q1 and Q2, utility captures diminishing throughput returns without
entering either reward. Let $x_{i,t}>0$ be the throughput of user $i$, with $i=1,\ldots,N_{\mathrm{NR}}+N_{\mathrm{WF}}$ and $x_{i,t}=\bar\Gamma_{k,t}$ for a user of RAT $k$ in this symmetric-within-RAT model. For positive throughput $x$, the utility $u(x)$ and its user average $\bar U_t$ are
\begin{equation}
 u(x)=\frac{\log(x/x_{\mathrm{L}})}{\log(x_{\mathrm{H}}/x_{\mathrm{L}})},
 \qquad
 \bar U_t=\frac{1}{N_{\mathrm{NR}}+N_{\mathrm{WF}}}\sum_i u(x_{i,t}),
 \label{eq:utility}
\end{equation}
where $\log$ denotes the natural logarithm and $x_{\mathrm{L}}$ and $x_{\mathrm{H}}$ are lower and upper throughput reference points. Following the concave-utility motivation in~\cite{Kelly1997ProportionalFairness,WangChan2013}, we use $(x_{\mathrm{L}},x_{\mathrm{H}})=(1,60)$ Mbps. Negative user utilities indicate throughput below $1$ Mbps.

\section{Fairness-Guided DQN-Based TXOP Control}
The implemented DQN observes the relative per-node throughput rather than
protocol-internal Wi-Fi states. Its scalar state is
$s_t=\bar\Gamma_{\mathrm{NR},t}/\bar\Gamma_{\mathrm{WF},t}$, evaluated for
$\bar\Gamma_{\mathrm{WF},t}>0$ without an additive numerical safeguard. This scalarization directly captures the relative service received by the two RATs and is invariant to a common scaling of both per-node throughputs. It therefore provides a compact state for fairness-oriented TXOP control. At the same time, different absolute-throughput pairs can produce the same ratio, so the state does not by itself encode total efficiency. Q2 additionally incorporates inter-RAT fairness and aggregate-throughput change into its reward. The action $a_t\in\{\mathrm{inc},\mathrm{dec},\mathrm{stay}\}$ increases, decreases, or preserves the TXOP:
\begin{equation}
 T_{t+1}=\clip\!\left(T_t\rho_t(a_t),T_{\min},T_{\max}\right).
 \label{eq:action}
\end{equation}
Here, $\rho_t(a_t)$ is the multiplicative action factor, with $\rho_t(\mathrm{inc})\sim\mathcal U(1,1.1)$, $\rho_t(\mathrm{dec})\sim\mathcal U(0.9,1)$, and $\rho_t(\mathrm{stay})=1$. The notation $\mathcal U(\alpha,\beta)$ denotes the continuous uniform distribution with lower and upper endpoints $\alpha\le\beta$. The lower and upper TXOP bounds are $(T_{\min},T_{\max})=(1,10^4)~\mu\mathrm{s}$. For a real input $v$, $\clip(v,T_{\min},T_{\max})=\min\{T_{\max},\max\{T_{\min},v\}\}$. Clipping precedes evaluation of the next state and reward.

% Reward timing is indexed by the resulting state: action a_t returns
% r_{t+1}. This is a notational alignment, not a changed reward function.
Reward design maps fairness preferences to control behavior. For $t\ge1$, let $r_t$ be the scalar reward returned after action $a_{t-1}$ using the resulting step-$t$ measurements. Its Q1 and Q2 versions are $r_t^{\mathrm{Q1}}$ and $r_t^{\mathrm{Q2}}$. Q1 rewards strict per-node balance:
\begin{equation}
 r_t^{\mathrm{Q1}}=
 \begin{cases}
 -1,&|s_t-1|>0.2,\\
 0.5,&0.1<|s_t-1|\le0.2,\\
 2,&|s_t-1|\le0.1.
 \end{cases}
 \label{eq:q1}
\end{equation}
Q2 uses fairness and the change from $\Gamma_{\mathrm{sum},t-1}$ to $\Gamma_{\mathrm{sum},t}$:
\begin{equation}
 r_t^{\mathrm{Q2}}=
 \begin{cases}
 1,&J_t^{\mathrm{RAT}}\ge0.8,\\
 -1,&J_t^{\mathrm{RAT}}<0.7,\\
 1,&0.7\le J_t^{\mathrm{RAT}}<0.8,\ 
       \Gamma_{\mathrm{sum},t}>\Gamma_{\mathrm{sum},t-1},\\
 0,&\text{otherwise}.
 \end{cases}
 \label{eq:q2}
\end{equation}
Thus, Q2 rewards all states with $J_t^{\mathrm{RAT}}\ge0.8$; throughput improvement is required only in the intermediate band $0.7\le J_t^{\mathrm{RAT}}<0.8$. In contrast, Q1 depends only on the distance of the throughput ratio from unity: ratios within 10\% receive the largest reward, deviations between 10\% and 20\% receive a smaller positive reward, and larger deviations are penalized. The two mappings consequently encode different operating preferences rather than different estimators of the same objective.

Let $\vect{\theta}$ be the online-network parameter vector and $Q(s,a;\vect{\theta})$ the action-value approximation for a scalar state $s$ and an action $a$. For a replayed transition $(s_t,a_t,r_{t+1},s_{t+1})$, the target and loss are
$y_t=r_{t+1}+\gamma\max_{a'}Q(s_{t+1},a';\vect{\theta}^{-})$ and
$\mathcal L(\vect{\theta})=\E[(y_t-Q(s_t,a_t;\vect{\theta}))^2]$,
respectively, where $\vect{\theta}^{-}$ denotes the target-network parameters
and the expectation is over replayed transitions~\cite{Mnih2015}. The network has a $1$--$128$--$64$--$3$ architecture, rectified linear unit (ReLU) hidden activations, and Adam optimization. Each network contains 8,707 parameters. Training uses experience replay and an $\epsilon$-greedy policy, where $\epsilon=0.1$ is the exploration probability without decay. Exploration is disabled during evaluation.

Two additional comparators complement the DQNs. Q2-u uses tabular Q-learning with the Jain index $J_t^u$ of the
NR-U and Wi-Fi per-node throughput utilities defined by~\eqref{eq:utility} as its state. The index is rounded to two decimal places for Q-table lookup. Its reward is $1$ for $J_t^{u}>0.8$, $-1$ for $J_t^{u}<0.7$, and $2$ for $0.7\le J_t^{u}\le0.8$ when the utility-transformed average throughput across all nodes increases; otherwise, it is $0$. In contrast, the default-referenced fixed-TXOP baseline is a non-learning comparator. For each priority class $p\in\{1,2,3,4\}$ and population, it selects the TXOP according to the following constrained optimization:
\begin{subequations}\label{eq:fixed}
\begin{align}
T_p^*
&=\arg\max_{T\in\mathcal T_p}\Gamma_{\mathrm{sum}}(T),
\label{eq:fixed-a}\\
\mathrm{s.t.}\quad
&\bar\Gamma_{\mathrm{WF}}(T)
\ge
\bar\Gamma_{\mathrm{WF}}(T_p^{\mathrm{def}}).
\label{eq:fixed-b}
\end{align}
\end{subequations}
Here, $T$ is a candidate TXOP, $\mathcal T_p$ is its deterministic
search set for priority class $p$, $T_p^*$ is the selected TXOP, and
$T_p^{\mathrm{def}}\in\mathcal T_p$ is the corresponding default TXOP.
The objective in~\eqref{eq:fixed-a} maximizes aggregate throughput,
whereas the constraint in~\eqref{eq:fixed-b} requires the per-node
Wi-Fi throughput to be no lower than that obtained under the default
TXOP for the same priority class and population. The functions $\Gamma_{\mathrm{sum}}(T)$ and $\bar\Gamma_{\mathrm{WF}}(T)$ are the analytical total and per-node Wi-Fi throughputs at candidate $T$, with priority and population held fixed. The sweep uses $1~\mu\mathrm{s}$ steps over $1$--$2000~\mu\mathrm{s}$ for Priority 1, $1$--$3000~\mu\mathrm{s}$ for Priority 2, and $1$--$8000~\mu\mathrm{s}$ for Priorities 3 and 4. No tuning seeds are required. Including $T_p^{\mathrm{def}}$ makes the constraint feasible.
Thus, the fixed baseline improves total throughput only over candidates that do not degrade Wi-Fi relative to default coexistence.

\subsection{Feasibility and Baseline Interpretation}
The default reference differs from a strict 3GPP reference-network test, which compares Wi-Fi service with an equal-sized all-Wi-Fi network~\cite{Kakkad2023TVT,Luo2022ICC}. In the implemented environment, the minimum NR-U channel-holding duration is
$T_{\mathrm{hold}}^{\min}=T_{\min}+D_{\mathrm{NR}}=501~\mu\mathrm{s}$.
Even at $T_{\min}$, NR-U nodes still contend and collide. Reducing TXOP therefore does not reproduce the all-Wi-Fi access process. If the highest attainable Wi-Fi throughput over the search set is below the strict reference, that constrained problem is infeasible. This is a limitation of TXOP-only control, not of the fairness criterion; joint adjustment of CW or defer duration may be necessary~\cite{Kakkad2023TVT,Luo2022ICC}.

The constraint in~\eqref{eq:fixed-b} instead protects Wi-Fi relative to default coexistence in the same configuration. It does not establish full 3GPP reference-network fairness. Q1 and Q2 likewise implement reward preferences rather than hard fairness guarantees. Q2-u provides a learned tabular comparator, whereas $T_p^*$ results from a deterministic constrained search for each configuration. These comparisons distinguish learning under an alternative fairness preference from configuration-specific non-learning TXOP optimization and characterize the associated tradeoffs without attributing all performance differences to the learning mechanism alone.

\begin{table*}[t]
\caption{Simulation and DQN Learning Configuration}
\label{tab:parameters}
\centering
\footnotesize
\setlength{\tabcolsep}{4pt}
\begin{tabularx}{\textwidth}{@{}p{0.18\textwidth}X p{0.18\textwidth}X@{}}
\toprule
\textbf{Parameter} & \textbf{Value} & \textbf{Parameter} & \textbf{Value}\\
\midrule
$(N_{\mathrm{NR}},N_{\mathrm{WF}})$ & $(5,5)$; $(2,8)$; $(8,2)$ &
Traffic; topology & Saturated; one collision domain\\
NR-U priorities & 1--4; Priority 3 for transfer &
NR-U initial CW & $(4,8,16,16)$ for Priorities 1--4\\
NR-U maximum stage & $(1,1,2,6)$ for Priorities 1--4 &
Wi-Fi initial CW; stage & $32$; $5$\\
Physical-layer rates & NR-U/Wi-Fi: $75.8/54$ Mbps &
Wi-Fi payload; idle slot & $2048$ bytes; $9~\mu\mathrm{s}$\\
$D_{\mathrm{NR}}$; TXOP bounds & $500~\mu\mathrm{s}$; $[1,10^4]~\mu\mathrm{s}$ &
DQN widths; optimizer & $1$--$128$--$64$--$3$; Adam\\
Learning rate; $\gamma$ & $10^{-3}$; $0.9$ &
Replay capacity & $10^4$ transitions\\
Minibatch size $N_{\mathrm b}$ & $64$ transitions &
Exploration; target update & $\epsilon=0.1$; every 100 steps\\
Training episodes & $1000$ &
Steps per episode & Training: at most 1000; evaluation: 1000\\
Q1 thresholds; rewards & $(0.2,0.1)$; $(-1,0.5,2)$ &
Q2 fairness thresholds & Lower/upper: $0.7/0.8$\\
Utility references & $(x_{\mathrm L},x_{\mathrm H})=(1,60)$ Mbps &
Independent trials & 20 predetermined reporting seeds\\
\bottomrule
\end{tabularx}
\end{table*}
The DQN-specific learning parameters in Table~\ref{tab:parameters}
apply to Q1 and Q2.

\section{Performance Evaluation}
\subsection{Protocol and Statistical Reporting}
Each result uses 20 predetermined reporting seeds with the same modeled configuration across methods. Each DQN is trained independently, then frozen and evaluated without exploration. For Q2-u, each reporting seed independently trains a Q-table. Evaluation freezes this table and selects a maximum-Q action from the current state to update TXOP at each step. The reported Q2-u metrics are time averages over the full evaluation trajectory rather than instantaneous values at the terminal TXOP. For Q1 and Q2, training episodes have at most 1000 steps; evaluation episodes have 1000 steps. Fixed TXOPs are selected by deterministic analytical sweeps independently of the reporting trials. No policy or threshold is tuned using reporting results.

For a reported metric $z$, let $\bar z$ and $s_z$ be its sample mean and sample standard deviation across the 20 trials. Error bars show $\bar z\pm t_{0.975,19}s_z/\sqrt{20}$, where $t_{0.975,19}$ is the $97.5$th percentile of a Student $t$ distribution with 19 degrees of freedom. The same reporting seeds are reused across methods so that environmental variation is aligned before pairwise comparison. For methods $A$ and $B$, let $d_h=z_h^A-z_h^B$ denote the paired difference
under reporting seed $h$. Its 95\% CI is
$\bar d\pm t_{0.975,19}s_d/\sqrt{20}$, where $\bar d$ and $s_d$ are the sample
mean and sample standard deviation of the 20 paired differences. Statistical improvement is claimed only when the paired CI excludes zero, rather than from marginal error-bar overlap.

\subsection{Computational Requirements}
Let $L$ be the number of parameterized fully connected layers and $n_j$ the width of layer $j\in\{0,\ldots,L\}$, where $j=0$ denotes the input. The implemented network has $L=3$ and $(n_0,n_1,n_2,n_3)=(1,128,64,3)$. A forward pass requires $\mathcal O(\sum_{j=1}^{L}n_{j-1}n_j)$ multiply--accumulate operations, where $\mathcal O$ denotes asymptotic order; for the implemented widths, this is 8,512 multiply--accumulate operations. A training update with minibatch size $N_{\mathrm b}$ scales this order by $N_{\mathrm b}$, in addition to replay storage. Fixed TXOP has constant online decision cost after $\mathcal O(|\mathcal T_p|)$ analytical evaluations, where $|\mathcal T_p|$ is the number of candidate TXOPs. With the $1~\mu\mathrm{s}$ sweep, $|\mathcal T_p|$ equals 2000, 3000, 8000, and 8000 for priority classes 1--4, respectively. The adapted bandit in~\cite{Bajracharya2023TVT} instead stores and updates per-arm statistics without a network forward pass. Q2-u uses tabular Q-values and therefore requires table lookup rather than neural network inference during evaluation. Hence, the fixed baseline concentrates cost offline, while the learned comparators differ in their online state representations and decision mechanisms.

The implemented online network contains only $8{,}707$ trainable parameters, while the target network has the same size and is used only during training. The controller observes only the scalar per-node throughput ratio, without channel-state information or protocol-internal Wi-Fi variables. This low-dimensional observation and modest inference cost support lightweight coexistence control.

\subsection{Priority-Class Tradeoff}
\begin{figure*}[t]
\centering
\includegraphics[width=2.\columnwidth]{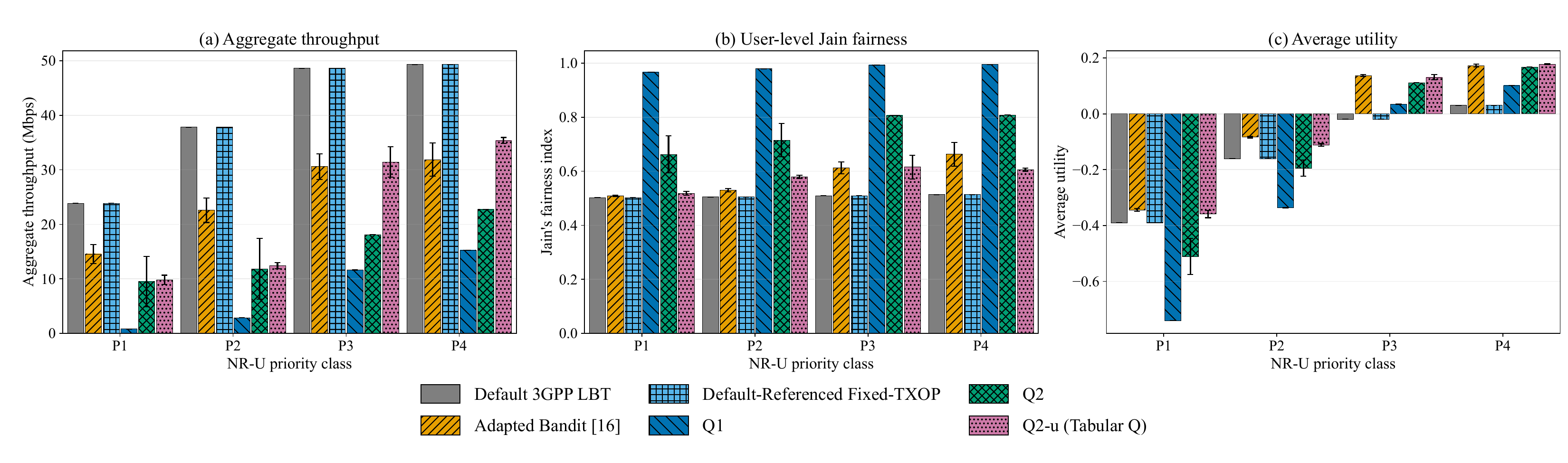}

\caption{Performance under $(N_{\mathrm{NR}},N_{\mathrm{WF}})=(5,5)$: (a) aggregate throughput, (b) user-level Jain fairness $J_t^{\mathrm{usr}}$, and (c) average utility. Error bars denote 95\% CIs over 20 trials. For Q1 and Q2, utility is evaluated only after training; Q2-u is the utility-fairness-based tabular Q-learning comparator.}
\label{fig:priority}
\end{figure*}

Priority classes change the contention process. Figure~\ref{fig:priority} compares total throughput, user-level fairness, and utility under $(N_{\mathrm{NR}},N_{\mathrm{WF}})=(5,5)$. Q1 achieves fairness close to unity, but its Priority-1 result also illustrates the low-throughput cost of balancing both RATs. Q2 recovers throughput by accepting a broader fairness region. Default LBT provides high aggregate throughput but substantially lower fairness. Under Priority 3, Q2 improves aggregate throughput over Q1 by $6.47$ Mbps (paired 95\% CI: $[6.45,6.48]$ Mbps). These observations support a tradeoff claim, not uniform superiority. The policy that most strongly equalizes per-node service need not maximize aggregate throughput or post-training utility, while the high-throughput default operating point can remain unattractive from a fairness perspective. Reporting the three metrics together therefore avoids interpreting any single objective as a complete coexistence criterion.

Q2-u provides the utility-fairness-based tabular comparison, whereas Q1 and Q2 employ the proposed DQN formulations. The default-referenced fixed-TXOP comparator solves~\eqref{eq:fixed} to assess configuration-specific offline tuning while protecting Wi-Fi relative to default coexistence. Under the tested priority classes, the constrained sweep selects the default operating point, so the resulting fixed-TXOP performance coincides with Default 3GPP LBT. This outcome indicates that, under the adopted Wi-Fi protection constraint, TXOP-only offline tuning does not improve aggregate throughput beyond the default configuration.

\subsection{Transfer to Asymmetric Populations}
\begin{figure*}[!t]
\centering
\includegraphics[width=2.0\columnwidth]{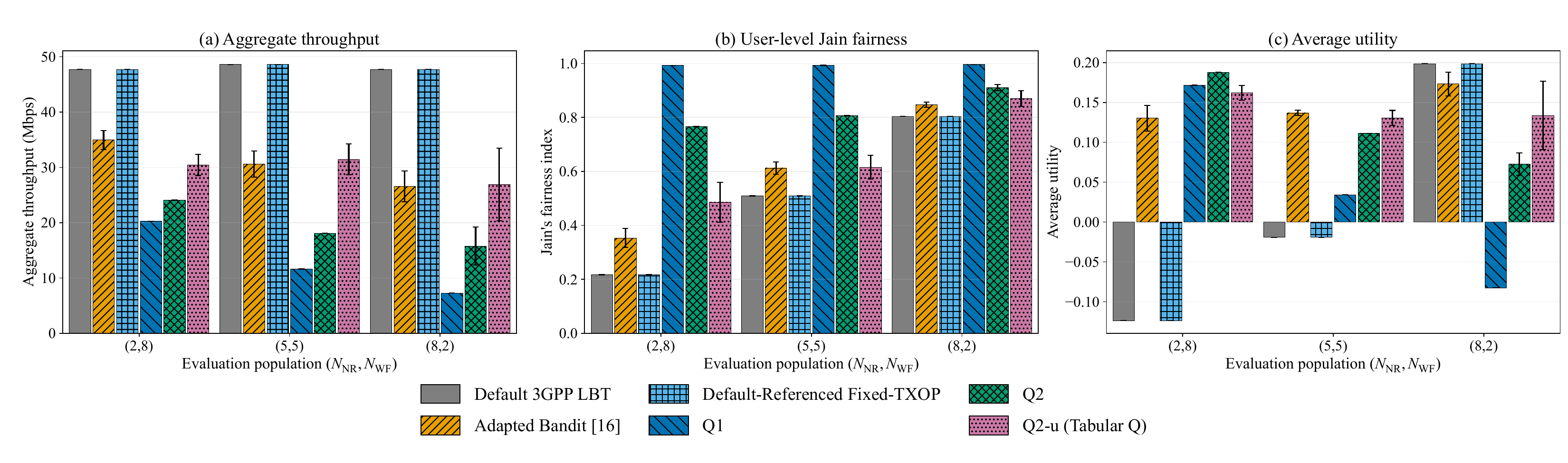}
\caption{Priority-3 transfer of Q1, Q2, and Q2-u trained under $(5,5)$ and evaluated without parameter updates under $(2,8)$, $(5,5)$, and $(8,2)$: (a) throughput, (b) user-level Jain fairness $J_t^{\mathrm{usr}}$, and (c) utility. The default-referenced fixed-TXOP baseline is recomputed for each population; statistical conventions follow Fig.~\ref{fig:priority}.}
\label{fig:asymmetry}
\end{figure*}

Frozen-policy evaluation tests transfer without retraining. For each reporting seed, the Priority-3 Q1/Q2 networks and Q2-u Q-table are trained under $(5,5)$ and reused without parameter updates under $(2,8)$, $(5,5)$, and $(8,2)$. Figure~\ref{fig:asymmetry} reports $J_t^{\mathrm{usr}}$ to account for unequal RAT populations. Q1 maintains high fairness but loses throughput as NR-U becomes dominant. Q2 provides a less restrictive operating point, and its utility under $(2,8)$ exceeds that of the utility-fairness-based Q2-u comparator; Q2-u is therefore not uniformly utility-superior across the tested populations. The default-referenced fixed-TXOP baseline is recomputed separately for each population; under the tested populations, the constrained solution again coincides with the corresponding default operating point. Under $(N_{\mathrm{NR}},N_{\mathrm{WF}})=(8,2)$, Q1 improves user-level Jain fairness over Q2 by $0.086$ (paired 95\% CI: $[0.074,0.098]$). This case also illustrates why $J_t^{\mathrm{usr}}$ is reported in addition to $J_t^{\mathrm{RAT}}$: under asymmetric populations, the two indices apply different population weightings and can therefore assign different fairness values unless the per-node throughputs of the two RATs are equal.

The transfer results also separate policy generalization from configuration-specific optimization. Q1 and Q2 retain the parameters learned under the symmetric population, whereas the fixed baseline is recomputed for each population. The observed degradation under population mismatch characterizes the limits of transfer across configurations. The scalar throughput ratio is compact, but different populations can produce similar ratios while corresponding to different absolute throughput levels and contention dynamics.

\subsection{Scope of Claims}
The model isolates TXOP control under saturated traffic and a single collision domain, excluding bursty queues, hidden terminals, spatial reuse, and multi-cell coupling. The scalar state also omits queue occupancy and protocol-internal Wi-Fi variables, which keeps inference lightweight but limits observability under non-saturated traffic. Strict reference-network fairness may require joint control of CW or sensing parameters~\cite{Kakkad2023TVT}. These limitations motivate traffic-aware, partially observable, and multi-parameter extensions.
 
\section{Conclusion}
This correspondence developed a fairness-guided DQN with a scalar throughput-ratio state for NR-U TXOP control. Q1 favors strict per-node balance, whereas Q2 accepts a broader fairness region. Priority-class and asymmetric-population results expose the fairness--throughput tradeoff and transfer limitations. The default-referenced fixed-TXOP baseline provides a configuration-specific offline comparator, while concave utility remains a post-training metric for Q1 and Q2. Q2-u additionally provides an independently trained utility-fairness-based tabular Q-learning comparison. Traffic-aware and multi-parameter coexistence are future extensions.

\bibliographystyle{IEEEtran}
\bibliography{References_revised}

\end{document}